# Which cities' paper output and citation impact are above expectation in information science? Some improvements of our previous mapping approaches

Lutz Bornmann [$] and Loet Leydesdorff [§]

[$] Max Planck Society, Administrative Headquarters, Hofgartenstr. 8, 80539 Munich, Germany; bornmann@gv.mpg.de (corresponding author).

[§] Amsterdam School of Communication Research, University of Amsterdam, Kloveniersburgwal 48, NL-1012 CX Amsterdam, The Netherlands; loet@leydesdorff.net.

**Abstract**
Bornmann and Leydesdorff (in press) proposed methods based on Web-of-Science data to identify field-specific excellence in cities where highly-cited papers were published more frequently than can be expected. Top performers in output are cities in which authors are located who publish a number of highly-cited papers that is statistically significantly higher than can be expected for these cities. Using papers published between 1989 and 2009 in information science improvements to the methods of Bornmann and Leydesdorff (in press) are presented and an alternative mapping approach based on the indicator $I3$ is introduced here. The $I3$ indicator was introduced by Leydesdorff and Bornmann (in press).

# 1    Introduction

The goal of our mapping approaches (Bornmann & Leydesdorff, in press; Bornmann, Leydesdorff, Walch-Solimena, & Ettl, in press; Bornmann & Waltman, in press) is to produce regional maps showing where excellent papers have emerged and where these papers have occurred frequently. Spatial bibliometrics has attracted a lot of attention: In general it pays off for the sciences within a country to identify and expand regional centers of excellence (with specific financial support). As a rule, there is a high probability of co-operation between scientists working at a short physical distance (Katz, 1994). In *Nature News* Van Noorden (2010) discussed urban regions producing the best research and whether their success could be replicated elsewhere. Living Science (http://www.livingscience.ethz.ch/), created by Luis Bettencourt (Los Alamos National Laboratory in New Mexico) and collaborators under Dirk Helbing at the ETH Zurich track in real time where preprints in *arXiv* are published.

Our most recent mapping approach (Bornmann & Leydesdorff, in press) considers the expected number in addition to the observed number of highly-cited papers for a city. If papers in the top-10% within a field, for example, are defined as the highly-cites papers, 10% of all papers published by authors located in a city provides the expected number. The observed number can be tested statistically against this expected number.

For example, if authors located in a city have published 1,000 papers, one would expect for statistical reasons that approximately 100 (that is, 10%) would also belong to the top-10% most-highly cited papers. An observed number of 70 highly-cited papers for this city may seem as a large number compared to other cities, but the specification of the expectation changes the appreciation. This approach has drawn considerable attention in science journalism (see, e.g., http://www.physorg.com/news/2011-03-european-team-scientific-relevance-city.html or http://blogs.discovermagazine.com/80beats/2011/03/22/the-best-cambridge-london-and-worst-moscow-taipei-cities-for-science/).

Using data from information science as an example we introduce here improvements and extensions of our approach. New programs and guidelines to generate maps have been made available at http://www.leydesdorff.net/topcity (for testing against the expected top-cited papers) and at http://www.leydesdorff.net/software/i3. On the latter page, a user can find all software needed for the mapping based on the Integrated Impact Indicator (*I3*). *I3* was introduced by Leydesdorff and Bornmann (in press). Using percentile ranks (e.g., top-1%, top-10%, etc.), *I3* integrates the citation curve over a publication set into an indicator value after normalization of the citation curves to the same scale.

## 2 Methods

### 2.1 Statistical procedure

*Comparison of the observed and expected top-cited paper numbers*

The $z$ test for two independent proportions (Sheskin, 2007, pp. 637-643) can be used for evaluating the degree to which an observed number of top-cited papers for a city differs from the value that would be expected on the basis of randomness in the selection of papers (Bornmann & Leydesdorff, in press). The value of $z$ is positively signed if the observed number of top papers is larger than the expected number and negatively signed in the reverse case. An absolute value of $z$ larger than 1.96 indicates statistical significance at the 5% level ($p<.05$) for the difference between observed and expected numbers of top-cited papers (marked with an asterisk *). Due to the large number of city tests being conducted (here: n=342), especially highly significant p values ($p<.01$) should be considered as significant and interpreted (marked with at least two asterisks **; analogously, *** will be used to indicate $p<.001$).

Using this statistical test, we designed the city circles which are visualized on the map using different colours and sizes. The radii of the circles are calculated by using: |observed value – expected value| + 1. The "+1" prevents the circles from disappearing if the observed

ratio is precisely equal to the expected one. Furthermore, the circles are coloured green if the observed values are larger than the expected values. We use dark green if both the expected value is at least five (and a statistical significance test is legitimate) and $z$ is statistically significant; light green indicates a positive, but statistically non-significant result. The in-between colour of lime green is used if the expected value is smaller than five and a statistical significance test hence should not be calculated. One should be cautious with interpretations of results below this threshold value.

In the reverse case that the observed values are smaller than the expected values the circles are coloured red or orange, respectively. They are red if the observed value is significantly smaller than the expected value and orange-red if the difference is statistically non-significant. If the requirement for the test of an expected value larger than five is not fulfilled the circle is coloured orange. If the expected value equals the observed value a circle is coloured grey.

*Calculation of I3*

One is inclined to conceptualize citation impact in terms of citations per publication, and thus as an average. The Impact Factor of journals (Garfield, 2006), for example, is an average. However, citation distributions are skewed and the average has the disadvantage that the number of publications is used in the denominator. Thus, a principal investigator has a higher average citation rate than s/he and her junior team together. However, the impact of the group is larger than that of the individual.

In other words, size matters for impact. Leydesdorff and Bornmann (in press) therefore replaced averaging with integration of the citation curve, but after qualifying the underlying publications in terms of their respective percentiles: a top-1% publication obtains 100 percent points whereas an average publication gets only 50 points. This rescaling from zero to hundred makes it possible to compare different sets and different citation distributions in terms of their impact.

The observed value of *I3* can be tested against an expected value that is proportional to the number of publications in the subset: $\exp(I3) = (n_i / \sum_i n_i) * \sum_i I3_i$. Because of the relatively large values of $I3 = \sum_i x_i * f(x_i)$, the difference between observed and expected can also be large.[1] Integrated impact is more easily significantly different than average impact.

Alternatively, one can test the impact/paper (*I3/n*) against $\sum_i I3_i / \sum_i n_i$. This is equivalent to testing the mean observed impact against the mean expected impact, analogously to the Relative Citation Rate (*RCR* = Mean Observed Citation Rate/ Mean Expected Citation Rate = *MOCR/MECR*) (Schubert & Braun, 1986) and will therefore be called *RI3R*. *RCR* and its derivatives such as *CPP/FCSm* (Moed, Debruin, & Van Leeuwen, 1995) and *NMRC* (Glänzel, Thijs, Schubert, & Debackere, 2009) have often been used for the normalization against a "world average." Unlike these authors, however, we do not divide the two means (Gingras & Larivière, 2011; Opthof & Leydesdorff, 2010), but use the expected value for the significance test.

As an average value *RI3R* (impact/paper) is sensitive to large values in the denominator and therefore small cities are relatively advantaged over larger ones. We use a minimum value of *N* = 5. The color scheme is similar as above, but the diameters of the nodes are based on the logarithm (*ln*) of the number of papers[2] involved because of the comparatively large differences between observed and expected values in the case of *I3*, and correspondingly small ones in the case of *I3/n*.

---

[1] The accumulation is not caused by (family-wise) repeated testing, but by aggregation. Therefore, Bonferroni correction is not appropriate.
[2] Because the logarithm of unity is zero and the node would thus disappear from the map, the value of *ln*(n+1) is used.

## 2.2 Procedure to generate the underlying data

*Data for the comparison of the observed and expected top-cited paper numbers*

The procedure to map the cities of the authors having published the top-cited papers in a certain field is described in detail in Bornmann and Leydesdorff (in press). In the following, we describe the most important steps and changes of the procedure. The top-10% of papers with the highest citation counts in a publication set can be considered as highly cited (Australian Research Council, 2011; Bornmann, Mutz, Marx, Schier, & Daniel, 2011). In this study we follow this classification and focus on the top-10% of papers published between 1989 and 2009 in information science, using a citation window for each paper from publication year up to the date of harvesting data from the Web of Science (WoS) 5.3 for this research (July 2011).

In a first step all papers with the document types "Article" were retrieved from the SSCI database which had been published between 1989 and 2009. To cover in this study the core journals of information science we included the same journals as used earlier by Leydesdorff and Persson (2010, p. 1623): *Annual Review of Information Science and Technology,*[3] *Information Processing & Management, Information Research, Journal of the American Society for Information, Science and Technology, Journal of Documentation, Journal of Informetrics, Journal of Information Science*. We restricted the search to articles (as document type) since the method proposed here is intended to identify excellence at the research front.

The search in WoS results in 6,242 papers which were saved as "full records" in packages of 500 articles each as plain text (e.g., savedrecs500.txt). The resulting 13 packages are then merged into a single file "data.txt" (see here the instructions on http://www.leydesdorff.net/software/isi/index.htm). This file is stored on the disk in a separate folder. The following procedure should be followed (see here also http://www.leydesdorff.net/topcity/index.htm). The programs cities1.exe and cities2.exe are copied from the website into the folder. These programs (including the respective user instructions) can be downloaded from http://www.leydesdorff.net/maps (see here Leydesdorff & Persson, 2010). The current version of cities1.exe no longer processes data downloaded from the WoS 4 interface, but only from the (new) WoS 5 interface.

---

[3] Since this journal publishes papers which are mostly classified by Thompson Reuters as "reviews" only a reduced number of papers could be included in this study.

Upon running, cities1.exe will prompt the user with the question: "Do you wish to skip the database management?" This question should the first time be answered with "N" (meaning: no). Thereafter, four questions follow: with the first and second questions one can set a threshold in terms of a minimal percentage of the total set of city-names in the data or set a minimum number of occurrences. The default answers to the questions ("0") can all be accepted. The third and fourth questions enable the user to obtain a cosine-normalized data matrix and to generate network data. Both questions can for our purpose be answered with "N" (meaning: no).

The program cities1.exe creates among other files the file named cities.txt. This file contains all city entries from data.txt, but organized so that this data can be "geo-coded," that is, provided with latitudes and longitudes on a map. The content of cities.txt can be copied-and-pasted into the GPS encoder at http://www.gpsvisualizer.com/geocoder/ (using Yahoo! as source for the geocoding). Geo-coding can also be done automatically using the Sci$^2$ Tool available at https://sci2.cns.iu.edu/user/download.php.[4] A third possible tool for geo-coding (used here) is the geocode command developed by Ozimek and Miles (2011) for the software Stata (www.stata.com).

After saving the (corrected) geocoding text file (e.g., "geo.txt") this data can serve as input to cities2.exe. If geo.txt contains all entries from cities.txt in the same order but with the additional geo data, the program cities2.exe (or inst2.exe, *mutatis mutandis*) can be used for matching these files. This program produces a number of output files in various formats within the same folder. In a final step, we proceed with the statistical procedure. topcity4.exe (available at http://www.leydesdorff.net/topcity/topcity4.exe) is an updated version of topcity2.exe introduced by Bornmann and Leydesdorff (in press).

---
[4] The Sci$^2$ Tool uses Yahoo! for the geo-coding.

The program topcity4.exe firstly asks to specify a percentile level. In this study, we used the top-10% of the most cited papers, and accordingly ten percent (the default) was entered. The user is further asked for the wished minimum size of the sample/ city. Only cities with the minimum paper number entered here are considered in the visualization (in this study: five papers). The last question for the minimum in the top-set gives the user the possibility to enter a threshold for considering of cities with at least zero, one or more papers among the top 10% (in this study, we use the default of zero papers).

The main further development of topcity4.exe against topcity2.exe concerns the possibility to include papers published in more than one year (here: 1989 to 2009). topcity4.exe considers papers with the same publication year as the reference set for computing the percentiles. Furthermore, we used Rousseau's (in press) suggestion to count the percentiles as the number of papers with a lower than or equal to citation rate divided by the total number of papers. The addition of the equal to sign ("≤" instead of "<") warrants that all articles including those in a reference set of less than hundred have a chance of reaching the top (100%) level. However, we decided not to distinguish for document types within publication years.

The file "ztest.txt" is one output file of topcity4.exe, and can be uploaded into the GPS Visualizer at http://www.gpsvisualizer.com/map_input?form=data. The data relevant for statistical analysis are provided in the file ucities.dbf which can be opened using Excel or SPSS. If more than a single co-author but with the same address is provided on a publication, this leads to a single city occurrence in the output. If the scientists are affiliated with departments in different cities, the different city names are used in the programs. The counting of occurrences in this study (so-called "integer counting") follows the procedure of how author addresses on publications are gathered by Thomson Reuters for inclusion in the WoS.

In "ztest.txt" the city entries from the WoS data are organized so that aggregated city occurrences can be visualized on a map, that is, provided with latitudes and longitudes (the source of the coordinates was in this case Google). The file "py.txt" contains for each publication year the number of papers belonging to the top papers (here: 10%) and the minimum number of citations for being a top paper (here: 10%).

The webpage of the GPS Visualizer offers a number of parameters which can be set to visualize the information contained in "ztest.txt." We suggest to change the following parameters: change (a) "waypoints" into "default;" (b) "colorize using this field" into "custom field" and choose "color" in this field; (c) "resize using this field" into "custom field" (d) in "custom resizing field" "n" is written and (e) at "Maximum radius" replace 16 with 30 (or 25). After processing the GPS data, the Google map is displayed first in a small frame, but this map is also available as full screen. The map shows the regional distribution of the authors of highly cited papers (cities with authors who published at least one excellent paper in the sample). The opacity of the background map can be adjusted and other layouts are also available in Google or Yahoo!. With the instruments visualized on the left side of the map one can zoom into the map. (Initially, the global map is shown.) Using the freely available API of Google, one can upload the html to one's own website.

For the maps presented here we zoomed in on some regions, like Europe and the USA. In order to determine the quotient of observed and expected numbers of excellent papers for a specific city, one can click on the respective city. The number is then displayed in the respective labels. We advise to check the maps against the original data at a number of random places before exhibiting it on the web. The user has all statistical data available in the file ucities.dbf.

*Data processing for the calculation of I3*

Since we used the same data set for generating *I3*-maps as for the maps comparing observed and expected top-cited papers, we present in the following only a general description of the procedure to produce maps. The website at http://www.leydesdorff.net/software/i3 provides the routines to compute *I3* for a set of papers downloaded from the WoS, version 5. First, the download can be organized in a relational database using the program ISI.exe. ISI.exe uses as input the download in the tagged format of the WoS which is available in the same folder and named "data.txt". The output is a set of databases (.dbf) which can be read using Excel or SPSS. For example, authors are organized into au.dbf and email addresses into em.dbf.

The resulting files can be used by isi2i3.exe as input. This program transforms core.dbf into i3core.dbf, au.dbf into i3au.dbf, and cs.dbf into i3cs.dbf. The program may take a while; in the case of large files, one can perhaps leave it overnight.The resulting files (e.g., i3core.dbf) are only different from the input files in a number of additional fields: the field i3f provides the value of *i3* normalized as percentiles in relation to the set under study ("the field"), and i3j is normalized at the level of each journal. Percentile values are normalized with reference to publication years and document types, and also Rousseau's (in press) correction is used.

Note that the use of i3f is only sensible if the set consists of all papers published in single field with comparable citation characteristics (Garfield, 1979; Leydesdorff & Bornmann, 2011; Moed, 2010). Analogously, r6f and r6j provide values for the six percentile ranks used by the NSF: top-1%, top-5%, top-10%, top-25%, top-50%, and bottom-50% (Bornmann & Mutz, 2011; National Science Board, 2010). isi2i3.exe furthermore generates a number of summary tables that one can use: i3so.dbf summarizes the data after aggregation at the journal level ("so" for source); i3cntry.dbf for aggregation at the country level. I3inst.dbf

and i3au.dbf allow for aggregation at the institutional level and author level, respectively, using pivot tables in Excel or "Aggregate cases" in SPSS. Note that the results addresses are "integer counted": each record is counted as one, whereas fractional counting would require an additional routine to attribute credit proportionally in the case of multi-authored papers.

i3cs.dbf can be used as input for the generation of overlays to Google Maps strictly analagous to the procedures described above for cs.dbf. Instead of cities1.exe and cities2.exe, one uses in this case i3cit1.exe and i3cit2.exe. Instead of inst1.exe and inst2.exe, one uses analogously i3inst1.exe and i3inst2.exe. i3cit2.exe and i3inst2.exe directly produce the various output files among which is ztest.txt.

## 3    Results

Figure 1 shows the location of authors in Europe having published highly-cited papers in information science and the deviations of the observed from the expected number of top-10% cited papers per location (the circle radii). Since the underlying data of a map from WoS (bibliographic data) and Google (and Yahoo!) are error-prone (Bornmann, et al., in press), we decided to visualize only cities (n=342) with an article output of at least five papers (see above). There is a danger for cities in the data with a small number of papers that they result from private addresses of researchers or erroneous address entries. The global map is made online available at http://www.leydesdorff.net/lis11/lis11.htm. If one clicks on a circle here, a frame opens showing the number of observed versus expected values for the respective city, as well as an asterisk indicating whether the difference between the values is statistically significant or not.

In Figure 1, for example, Budapest is indicated by a very large dark green circle—one of the largest green circles in Europe—because of an observed value much larger than expected. In Budapest the well-known Information Science and Scientometrics Research Unit (ISSRU) is located at the Library of the Hungarian Academy of Sciences. In the description in the frame, the large and statistically highly significant difference between the observed ($n_o$=46) and the expected value ($n_e$=17.9) of top-10% highly-cited papers can be retrieved. (It follows that $N$ = 179 for "Budapest.")

Further large green circles on the map with a statistically highly significant difference ($p<.01$) between observed and expected values are visible for Zurich ($n_o$=27, $n_e$=5.6) in Switzerland (here the Professorship for Social Psychology and Research on Higher Education at the ETH Zurich is located), Amsterdam ($n_o$=34, $n_e$=13.0) (here the Amsterdam School of Communications Research, ASCoR, at the University of Amsterdam is located) and Leiden ($n_o$=36, $n_e$=10.8) (here the Centre for Science and Technology Studies, CWTS, is located) at Leiden University in the Netherlands as well as Wolverhampton ($n_o$=22, $n_e$=8.5) in England (here the School of Technology, University of Wolverhampton, is located).

Figure 2 shows the corresponding map focusing on the USA. There is only one circle visible with a statistically highly significant difference between the observed ($n_o$=24) and the expected value ($n_e$=9.9) for Philadephpia, PA (here both the College of Information Science and Technology at Drexel University and the Institute of Scientific Information—ISI; nowadays integrated into Thomson Reuters—are located). The other dark green circles on the map indicate cities with a statistically significant difference only at the 5% level. Extending the view to a global one (see here http://www.leydesdorff.net/lis11/lis11.htm), our results point to the fact that in addition to the described cities in Europe and the United States only one more city shows a highly significantly positive difference of observed against expected citation counts of the articles: Montreal in Canada: ($n_o$=23, $n_e$=9.2).

Figure 3 shows a zoom of Figure 1, but using the indicator *I3*. Different from Figure 1, many more cities are visible because the focus is no longer on those cities which contribute at the top-10% layer. (We suppressed city-nodes with less than five papers in both maps.) Major centers of activity such as Copenhagen, Sussex (SPRU), and Paris (Callon's group) are now also colored green. Several centers in the UK use information science and publish in the journals, but their contributions are below expectation in terms of citation impact (e.g., London, Edinburgh, and Cambridge). As noted, the *z*-test is more sensitive for differences between observed and expected values in the case of *I3* because of the possibly magnifying effect of the summation (integration) on this difference.

Figure 4 shows as the results of *z*-testing the *I3*-values for East Asia. We highlighted the impact far above expectation of Dalian. Furthermore, the absence of green-colored nodes in Japan is noticeable. Figure 5 confirms the impression of Figure 3 that this indicator provides a richer and more informed map about the situation in Northern America. The impact contribution of a major center as Bloomington Indiana, for example, is indicated as highly significant ($p<0.001$). Smaller centers of activity such as Baton Rouge in Louisiana are also colored green albeit with somewhat less significance. Centers on the Westcoast such as Stanford University and UCLA are also indicated.

Figure 6 finally shows the effect of using impact per paper. The major centers (such as Budapest, Leiden, and Wolverhampton) are no longer significant in terms of their impact. Smaller centers with a few high-quality papers during these two decades are foregrounded. Similarly in the USA, one can see (at http://www.leydesdorff.net/lis11/lis11ri3r.htm) that cities with large concentrations of papers such as Philadelphia lose their significance in terms of average impact whereas the 13 papers published with an address in Albuquerque NM are now indicated as on average highly significantly above expectation in terms of their impact.

# 4    Discussion

In this paper we have presented extensions and new rudiments of our mapping approach published recently (Bornmann & Leydesdorff, in press). The most important improvement in the comparison of the observed and expected top-10% highly cited papers for a city is the normalization for different publication years in the program topcity4.exe. This leads to a more flexible use of the approach. The new possibility to generate maps on the base of the recently introduced *I3*-metric allows for the visualization of an indicator for cities which considers both, productivity and normalized impact (similar to the *h* index). The *I3*-metric provides also the potential to compare observed and expected values.

The mapping results for information science show that we arrive at somewhat different results by using the top-10%-approach in comparison to the *I3*-approach. On the *I3*-maps many more cities are visible because the focus is no longer on the excellent papers only but all papers are included in the comparison. On the one side, this results in the visualization of major centers of activity (e.g., Copenhagen, Sussex, and Paris) which are not so prominent visible by using only the top-cited-approach. On the other side, several centers that use information science and publish in its journals have contributions below the expectation.

The different results of both mapping approaches point to the desirability to visualize the same data set including publication and citation numbers for cities using these different approaches. Only in this way it is possible to see (1) results in agreement which is an indication of reliability and (2) different results which can point to unreliable results (because they are somehow dependent on the method).

Both approaches, for example, show agreement about a spatial concentration of excellent scientific activity in Belgium and the Netherlands. Since both approaches point this out it seems to be a reliable result for information science. Bornmann, et al. (in press) propose to name such spatial concentrations of activity the "reverse N-effect." The formulation of the N-effect goes back to Garcia and Tor (2009). The reverse N-effect can then be defined as follows: "More competitors (here: prolific scientists) working within the same region produce better results … the better result may consist of a higher output of highly-cited papers" (Bornmann, et al., in press). Both Belgium and the Netherlands are characterized by some excellent research groups or institutes, respectively, which undertake information science research and research in related areas on a very high level.

There are several problems inherent to the mapping approaches proposed here. Bornmann, et al. (in press) formulate a comprehensive list of problems of which one should be aware. The two most important ones are the following:

1) There are circles on the maps that are not at the correct position. In the various routines, we try to avoid these misallocations, but misspellings, for example, may occur. The misallocations do have different sources: errors in the WoS data or erroneous coordinates provided by the geocoding.

2) High numbers of publications visualized on the map for one city might be due to the two following effects: (a) Many scientists located in this city (i.e., scientists at different institutions or departments within one institution) produced at least one excellent paper or (b) one or only a few scientists located in this city produced many influential papers. Assuming cities as units of analysis, one is not able to distinguish between these two configurations.

The maps produced by the approaches introduced here should always be checked carefully. Bornmann, et al. (in press) describe some advanced techniques to do this. Since the bibliometric data from the databases (WoS) and the geocodes for cities are error-prone, maps are never without any errors. This fact should always be considered in spatial bibliometrics (as an observer as well as a producer of maps). If the reader of this article finds some errors on the maps produced for this paper we appreciate a corresponding feedback.

# References

Australian Research Council. (2011). ERA 2010: citation benchmark methodology. Canberra, Australia: Australian Research Council.

Bornmann, L., & Leydesdorff, L. (in press). Which cities produce more excellent papers than can be expected? A new mapping approach—using Google Maps—based on statistical significance testing. *Journal of the American Society of Information Science and Technology*.

Bornmann, L., Leydesdorff, L., Walch-Solimena, C., & Ettl, C. (in press). How to map excellence in the sciences? A mapping approach made possible by using Scopus data. *Journal of Informetrics*.

Bornmann, L., & Mutz, R. (2011). Further steps towards an ideal method of measuring citation performance: the avoidance of citation (ratio) averages in field-normalization. *Journal of Inormetrics, 5*(1), 228-230.

Bornmann, L., Mutz, R., Marx, W., Schier, H., & Daniel, H.-D. (2011). A multilevel modelling approach to investigating the predictive validity of editorial decisions: do the editors of a high-profile journal select manuscripts that are highly cited after publication? *Journal of the Royal Statistical Society - Series A (Statistics in Society), 174*(4). doi: 10.1111/j.1467-985X.2011.00689.x.

Bornmann, L., & Waltman, L. (in press). The detection of "hot regions" in the geography of science: a visualization approach by using density maps. *Journal of Informetrics*.

Garcia, S. M., & Tor, A. (2009). The n-effect: more competitors, less competition. *Psychological Science, 20*(7), 871-877.

Garfield, E. (1979). Is citation analysis a legitimate evaluation tool? *Scientometrics, 1*(4), 359-375.

Garfield, E. (2006). The history and meaning of the Journal Impact Factor. *Journal of the American Medical Association, 295*(1), 90-93.

Gingras, Y., & Larivière, V. (2011). There are neither "king" nor "crown" in scientometrics: comments on a supposed "alternative" method of normalization. *Journal of Informetrics, 5*(1), 226-227. doi: 10.1016/j.joi.2010.10.005.

Glänzel, W., Thijs, B., Schubert, A., & Debackere, K. (2009). Subfield-specific normalized relative indicators and a new generation of relational charts: methodological foundations illustrated on the assessment of institutional research performance. *Scientometrics, 78*(1), 165-188.

Katz, J. S. (1994). Geographical proximity and scientific collaboration. *Scientometrics, 31*(1), 31-43.

Leydesdorff, L., & Bornmann, L. (2011). How fractional counting of citations affects the Impact Factor: normalization in terms of differences in citation potentials among fields of science. *Journal of the American Society for Information Science and Technology, 62*(2), 217-229. doi: 10.1002/asi.21450.

Leydesdorff, L., & Bornmann, L. (in press). Integrated Impact Indicators (I3) compared with Impact Factors (IFs): an alternative research design with policy implications. *Journal of the American Society of Information Science and Technology*.

Leydesdorff, L., & Persson, O. (2010). Mapping the geography of science: distribution patterns and networks of relations among cities and institutes. *Journal of the American Society for Information Science and Technology, 61*(8), 1622-1634. doi: 10.1002/Asi.21347.

Moed, H. F. (2010). Measuring contextual citation impact of scientific journals. *Journal of Informetrics, 4*(3), 265-277. doi: 10.1016/j.joi.2010.01.002.

Moed, H. F., Debruin, R. E., & Van Leeuwen, T. N. (1995). New bibliometric tools for the assessment of national research performance - database description, overview of indicators and first applications. *Scientometrics, 33*(3), 381-422.

National Science Board. (2010). Science and engineering indicators 2010. Arlington, VA, USA: National Science Foundation (NSB 10-01).

Opthof, T., & Leydesdorff, L. (2010). Caveats for the journal and field normalizations in the CWTS ("Leiden") evaluations of research performance. *Journal of Informetrics, 4*(3), 423-430.

Ozimek, A., & Miles, D. (2011). Stata utilities for geocoding and generating travel time and travel distance information. *Stata Journal, 11*(1), 106-119.

Rousseau, R. (in press). Percentile rank scores are congruous indicators of relative performance, or aren't they? *Journal of the American Society for Information Science and Technology*.

Schubert, A., & Braun, T. (1986). Relative indicators and relational charts for comparative assessment of publication output and citation impact. *Scientometrics, 9*(5-6), 281-291.

Sheskin, D. (2007). *Handbook of parametric and nonparametric statistical procedures* (4th ed.). Boca Raton, FL, USA: Chapman & Hall/CRC.

van Noorden, R. (2010). Cities: building the best cities for science. Which urban regions produce the best research - and can their success be replicated? *Nature, 467*, 906-908. doi: 10.1038/467906a.

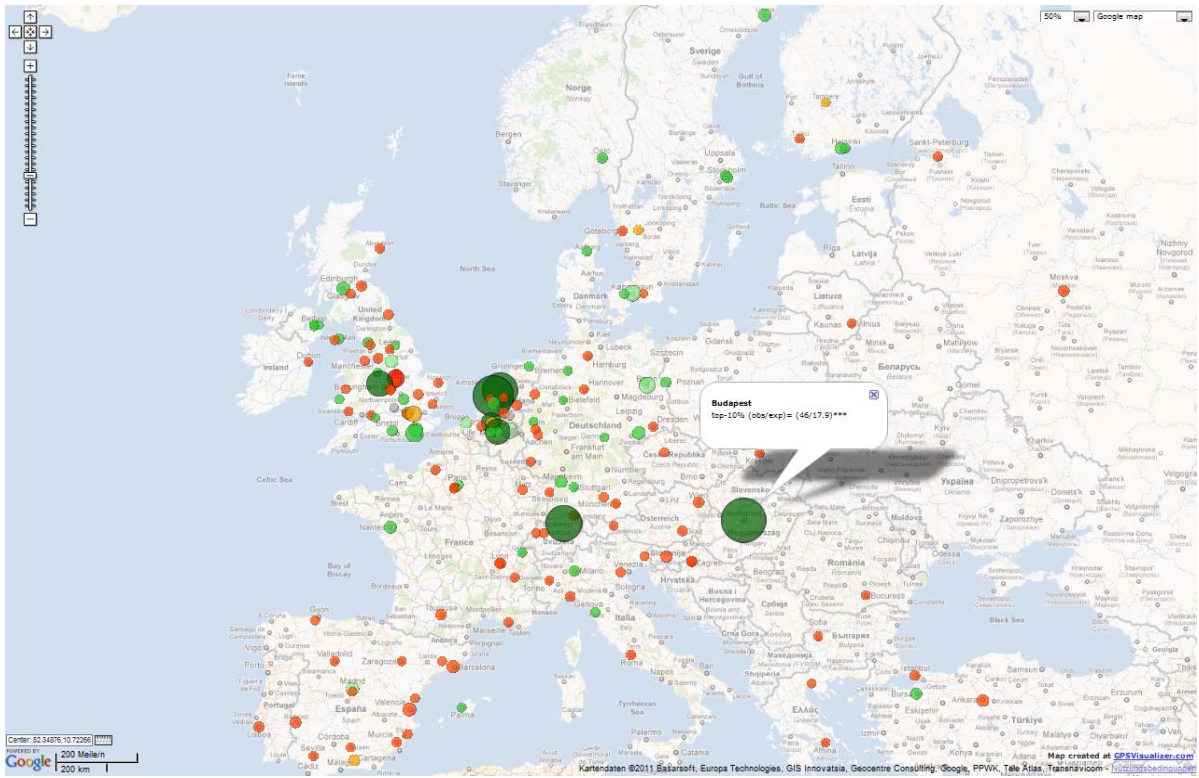

Figure 1. Cities in Europe with (top-10%) highly cited articles in information science during 1989 and 2009 (only cities are visualized with a total article output of at least five; see for the full map at http://www.leydesdorff.net/lis11/lis11.htm ).

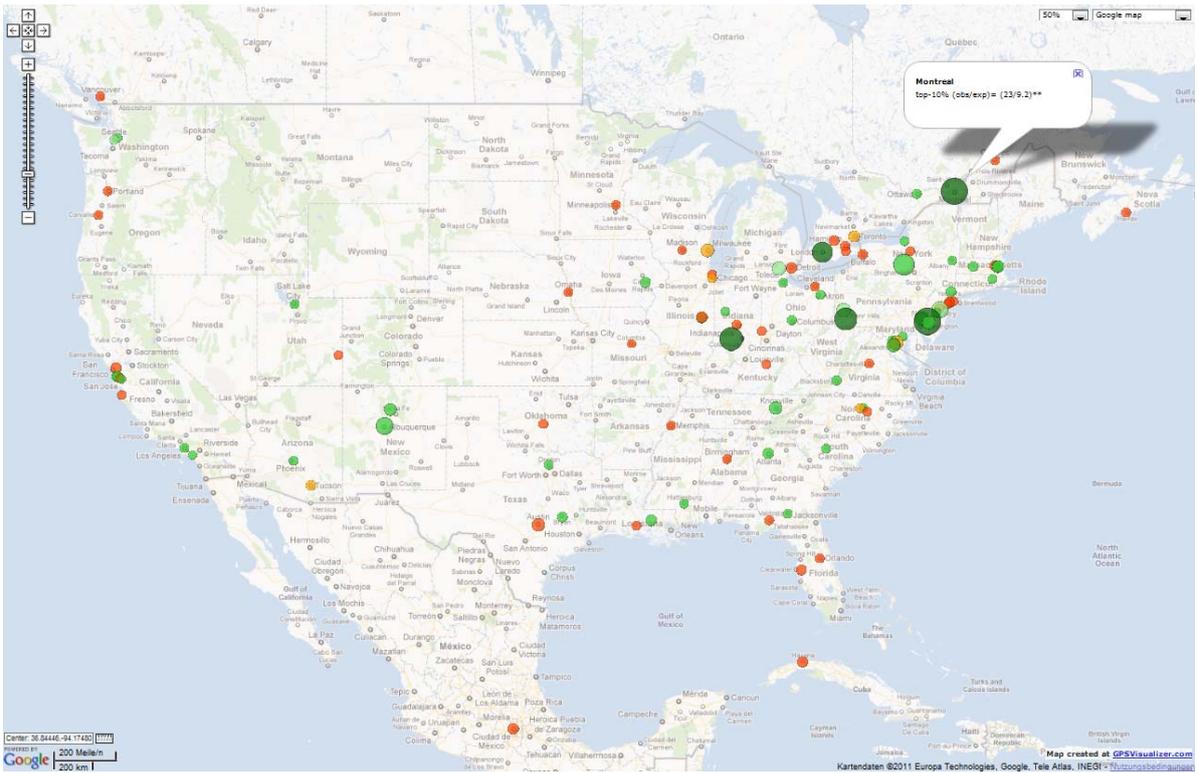

Figure 2. Cities in the USA with (top-10%) highly cited articles in information science during 1989 and 2009 (only cities are visualized with a total article output of at least five; see for the full map at http://www.leydesdorff.net/lis11/lis11.htm ).

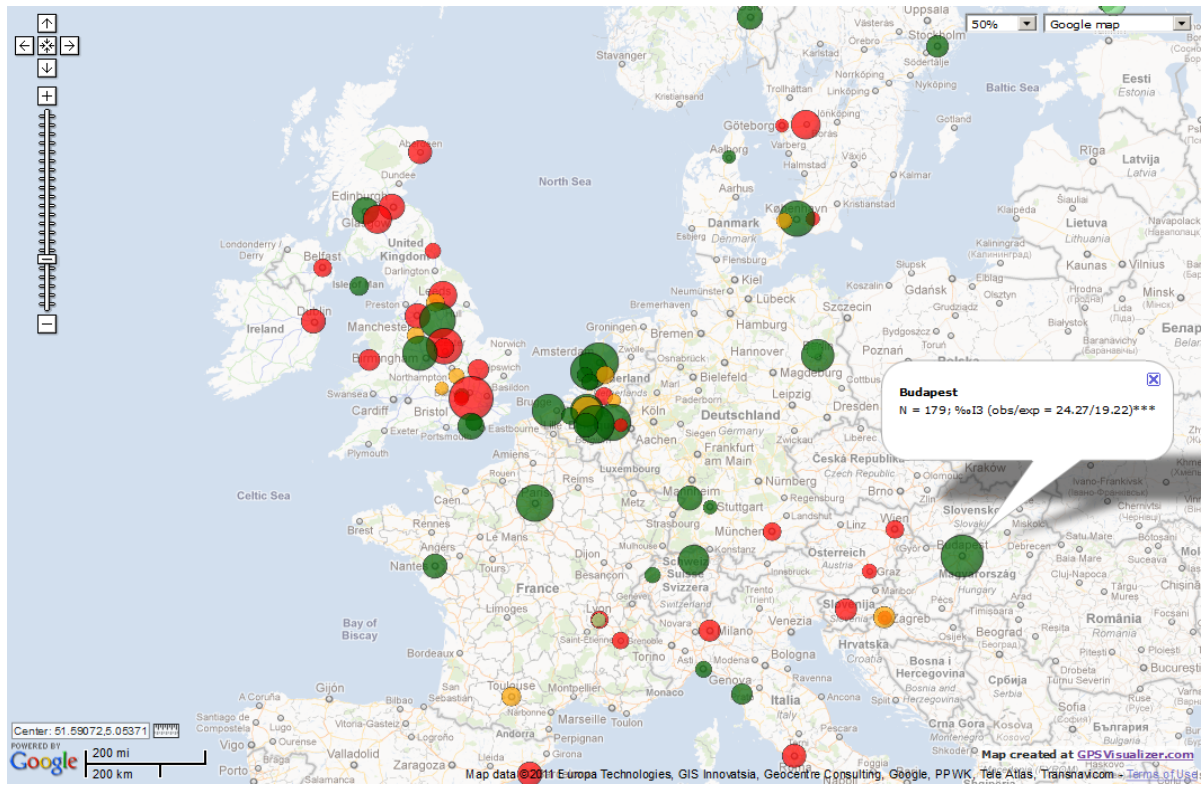

Figure 3. Integrated impact values for cities in Europe (information science, cities are visualized with a total article output of at least five during 1989 and 2009; see for the full map at http://www.leydesdorff.net/lis11/lis11i3.htm ).

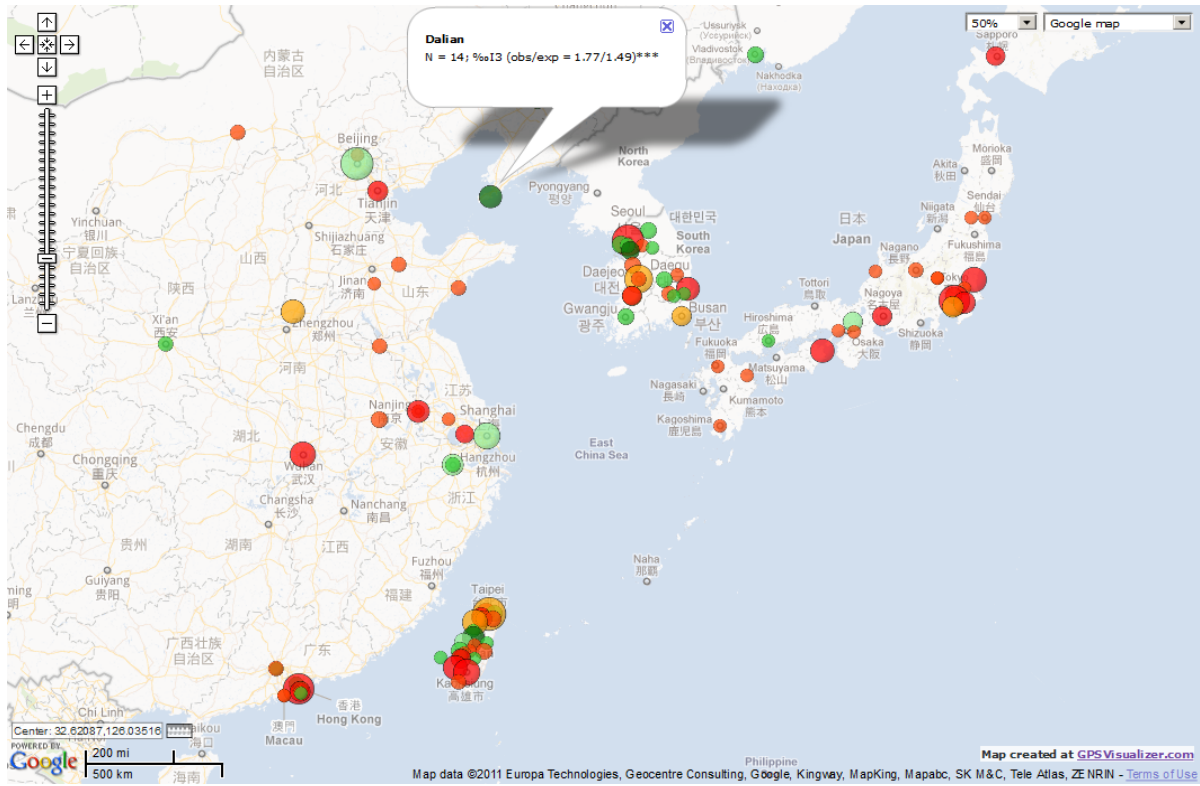

Figure 4. Integrated impact values for cities in East Asia (information science, cities are visualized with a total article output of at least five during 1989 and 2009; see for the full map at http://www.leydesdorff.net/lis11/lis11i3.htm ).

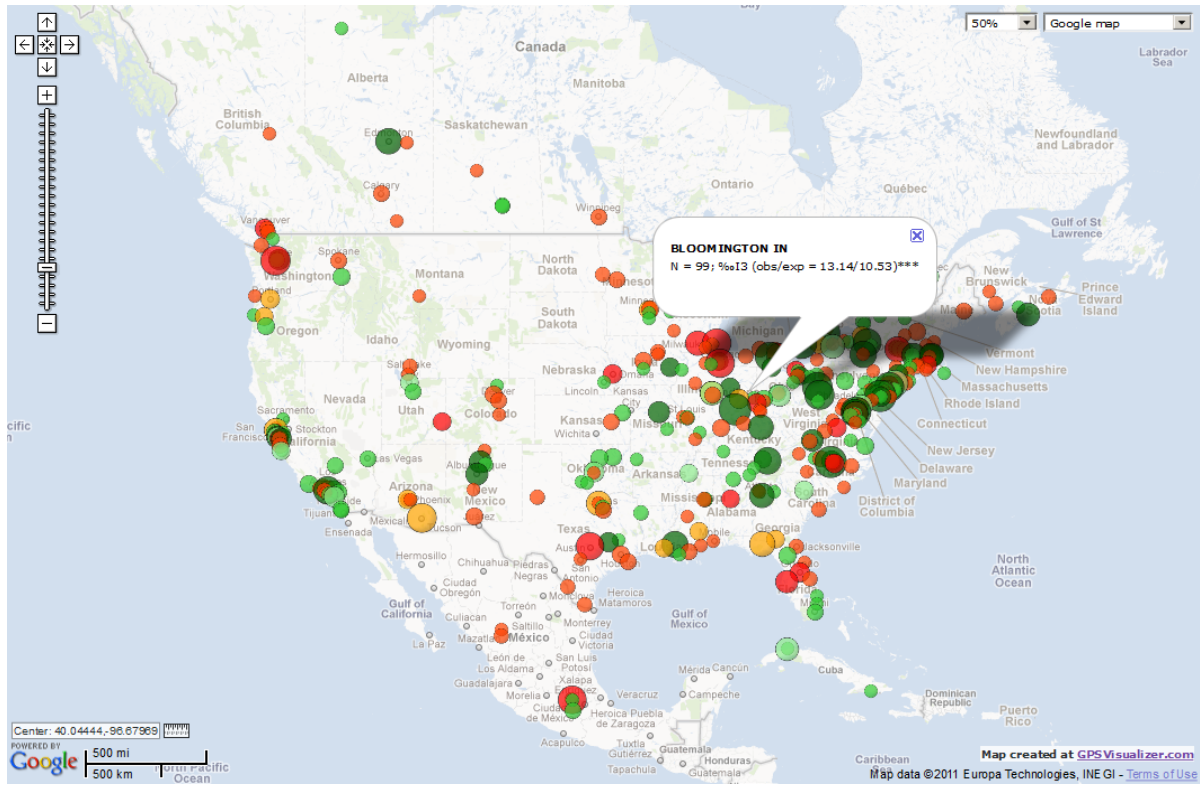

Figure 5. Integrated impact values for cities in Northern America (information science, cities are visualized with a total article output of at least five during 1989 and 2009; see for the full map at http://www.leydesdorff.net/lis11/lis11i3.htm ).

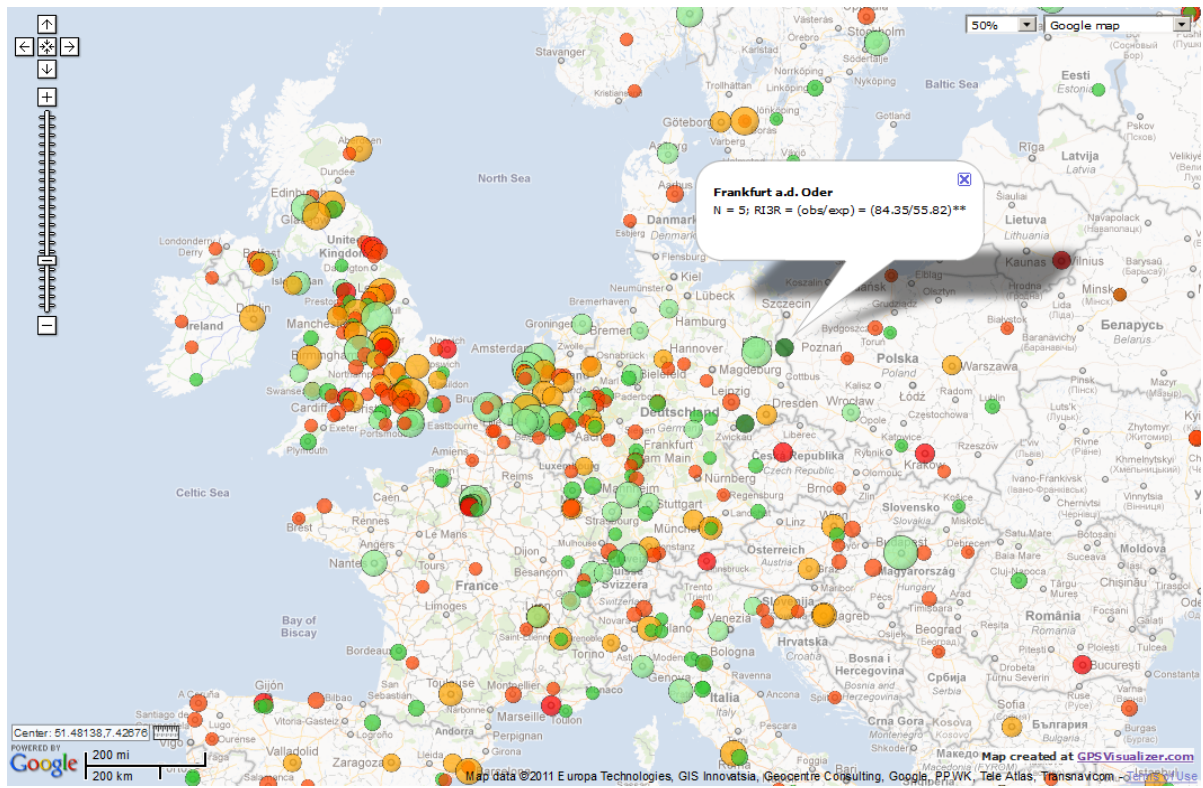

Figure 6. Integrated impact values (impact per paper) for cities in Europe (information science, cities are visualized with a total article output of at least 5 during 1989 and 2009; see for the full map at http://www.leydesdorff.net/lis11/lis11ri3r.htm )